\documentclass[12pt]{iopart}
\usepackage{iopams}  

\usepackage{amssymb}
\usepackage[mathscr]{euscript}
\usepackage{cite}

\newcommand{\bb}[1]{{\mathbb #1}}
\newcommand{\mc}[1]{{\mathcal #1}}
\newcommand{\mms}[1]{{\mathscr #1}}

\newcommand{\upbar}[1]{\,\overline{\! #1}}

\newcommand{\<}{\langle}
\renewcommand{\>}{\rangle}

\renewcommand{\phi}{\varphi}

\newcommand{\eqref}[1]{(\ref{#1})}

\begin{document}

\title[Lagrangian phase transitions]
{Lagrangian phase transitions in nonequilibrium thermodynamic systems}

\author{Lorenzo Bertini}
\address{
Dipartimento di Matematica, Universit\`a di Roma La Sapienza }
\ead{bertini@mat.uniroma1.it}

\author{Alberto De Sole}
\address{
Dipartimento di Matematica, Universit\`a di Roma La Sapienza }
\ead{desole@mat.uniroma1.it}

\author{Davide Gabrielli}
\address{
Dipartimento di Matematica, Universit\`a dell'Aquila}
\ead{gabriell@univaq.it}

\author{Giovanni Jona-Lasinio}
\address{
Dipartimento di Fisica and INFN, Universit\`a di Roma La Sapienza}
\ead{gianni.jona@roma1.infn.it}

\author{Claudio Landim} 
\address{
IMPA {\normalfont and}  CNRS UMR 6085, Universit\'e de Rouen} 
\ead{landim@impa.br}

\begin{abstract}
  In previous papers we have introduced a natural nonequilibrium free
  energy by considering the functional describing the large
  fluctuations of stationary nonequilibrium states. While in
  equilibrium this functional is always convex, in nonequilibrium this
  is not necessarily the case. We show that in nonequilibrium a new
  type of singularities can appear that are interpreted as phase
  transitions.  In particular, this phenomenon occurs for the
  one-dimensional boundary driven weakly asymmetric exclusion process
  when the drift due to the external field is opposite to the one due
  to the external reservoirs, and strong enough.
\end{abstract}


\vspace{2pc}

\maketitle

\section{Introduction}

Irreversible nonequilibrium phenomena have been central in statistical
mechanics research in the last decades. In the last ten years the
authors have developed a new approach to nonequilibrium statistical
mechanics inspired and supported by the analysis of stochastic lattice
gases \cite{BDGJL1a,BDGJL1b,BDGJL1c,BDGJL1d}.
This theory is applicable to a wide class of thermodynamic systems
where diffusion is the dominant mechanism.  
For example, as shown in \cite{Dcur1,Dcur2}, this theory leads to the
prediction of universality properties for current fluctuations.
A basic ingredient of the theory is the so called quasi-potential, a
concept introduced in the analysis of stochastically perturbed
dynamical systems \cite{FW}, which provides a natural definition of a
nonequilibrium thermodynamic potential.

In this letter we discuss the occurrence of singularities of the
quasi-potential for nonequilibrium systems with infinitely many
degrees of freedom.  
We analyze in detail the weakly asymmetric exclusion process and show
analytically that, when the external field is strong, 
these singularities do appear.  
The singularities of the quasi-potential are interpreted as
nonequilibrium phase transitions. Examples of phenomena of this kind
in a finite dimensional setting have been discussed in the literature
and have also been observed in simulations \cite{J1,J2,J3}. 
The present work is the first example in which a thermodynamic model,
that is a system with infinitely many degrees of freedom, is shown to
exhibit such a singular behavior.

\section{Macroscopic fluctuation theory}

The dynamical macroscopic behavior of the system in a $d$-dimensional
volume $\Lambda$ is described by a nonlinear driven diffusion type equation
of the form
\begin{equation}
\label{f1}
\rho_t + \nabla \cdot \sigma (\rho) E 
= \nabla \cdot D(\rho) \nabla \rho \,,
\end{equation}
where $\rho=\rho(t, x)$ represents the thermodynamic variable, e.g.\ the
density, and $\rho_t$ is its time derivative. 
The diffusion coefficient $D$ and the mobility $\sigma$ are
$d\times d$ matrices and $E$ denotes the external field.  The
transport coefficients $D$ and $\sigma$ satisfy the local Einstein
relation $D(\rho) = \sigma(\rho) \, s''(\rho)$ where $s$ is the
equilibrium free energy of the homogeneous system.  Equation
\eqref{f1} has to be supplemented by the appropriate boundary
conditions due to the interaction with the external reservoirs.  We
denote by $\bar\rho=\bar\rho(x)$ the stationary solution of
\eqref{f1}. 

The hydrodynamic equation \eqref{f1} can be derived from an underlying
microscopic dynamics through a suitable scaling limit. It
represents the typical behavior, as the number $N$ of degrees of
freedom diverges, of the empirical density profile $\rho_N(t,x)$
defined as the average number of particles at time $t$ in a
macroscopic infinitesimal volume around $x$. The validity of the
local Einstein relationship can then be deduced from the local microscopic
detailed balance \cite{S}.

The probability that in the time interval $[T_1, T_2]$ the evolution
of the variable $\rho_N$ deviates from the solution of the hydrodynamic
equation and is close to some trajectory $\rho$, is exponentially small
and of the form
\begin{equation}
\label{LD}
P\left( \rho_N(t,x) \approx \rho(t,x) \right) \;\approx\;
e^{-N I_{[T_1,T_2]} (\rho) }\,,
\end{equation}
where $I_{[T_1,T_2]} (\rho)$ is a functional which vanishes if $\rho$
is a solution of \eqref{f1}. The functional $I_{[T_1,T_2]}(\rho)$
represents the energetic cost necessary for the system to follow the
trajectory $\rho$.
In the case of stochastic lattice gases, the expression of
$I_{[T_1,T_2]}$ can be obtained from the microscopic dynamics as the
large deviation rate functional  \cite{BDGJL1a,BDGJL1b,BDGJL1c,BDGJL1d,dd}.
Following \cite{BDGJL3}, we next sketch a purely macroscopic argument
which yields the same conclusion.
Consider a time dependent variation $F= F(t,x)$ of the external field
so that the total applied field is $E+F$. Denote by $\rho^F$ the
corresponding solution of \eqref{f1}.  By minimizing the energy
dissipated by the field $F$ with the constraint that $\rho^F$ equals
the prescribed path $\rho$ we obtain that
\begin{equation}
\label{2.5}
I_{[{T_1},{T_2}]}(\rho)
= \frac 14 \int_{{T_1}}^{{T_2}} 
\big\langle F  \cdot \sigma(\rho) F \big\rangle \, dt \,,
\end{equation}
where $\langle \cdot \rangle$ is the integration over space and the
optimal field is given by $F=2\nabla H$, where $H$ is the unique
solution to the Poisson equation
\begin{equation}
\label{pois}
- 2 \nabla \cdot \sigma(\rho) \nabla H 
= \rho_t  + \nabla \cdot \sigma(\rho) E - \nabla \cdot D(\rho) \nabla \rho  
\end{equation}
which vanishes at the boundary of $\Lambda$ for any $t\in[T_1,T_2]$. 

The quasi-potential $V(\rho)$ is defined as the minimal cost to
reach the density profile $\rho$ starting from the stationary
profile $\bar\rho$:
\begin{equation}
\label{2.6}
V (\rho) =
  \inf \Big\{I_{{(-\infty,0]}}(\hat \rho)\,,\,\,
   \hat \rho \, : \: \hat \rho ({-\infty}) = \bar\rho \,,\,\, \hat
  \rho(0) =\rho \Big\}\,.
\end{equation}
Therefore, while $I_{[T_1,T_2]}(\hat\rho)$ measures how much a path
$\hat\rho$ is close to the solution of \eqref{f1}, the quasi-potential
$V (\rho)$ measures how much a profile $\rho$ is close to the
stationary solution $\upbar{\rho}$. Moreover, $V$ is proportional to
the total work done by the external field along the
optimal time evolution to reach the density profile $\rho$
\cite{BDGJL3}. In the context of nonequilibrium stationary states of
stochastic lattice gases the quasi-potential gives the asymptotics, as
the number of degrees of freedom diverges, of the probability of
observing a static fluctuation of the density:
$\mu\left( \rho_N(x) \approx \rho(x) \right) \;\approx\;
e^{-N V (\rho) }$,
where $\mu$ is the stationary state of the microscopic dynamics.  This
makes natural to interpret $V$ as a nonequilibrium free energy. 
In particular, for equilibrium systems $\mu$ has the standard Gibbs form and the
quasi-potential coincides with the variation of the free energy.

\section{Hamiltonian picture}

By considering the functional $I_{[T_1, T_2]}$ defined in \eqref{2.5}
as an action functional, i.e.\ $I_{[T_1, T_2]}(\rho)=\int_{T_1}^{T_2} \bb
L(\rho,\rho_t) \, dt$ for a Lagrangian $\bb L(\rho,\rho_t)$ obtained
by solving \eqref{pois} and expressing the external field $F$ in terms
of $\rho$ and $\rho_t$, the variational problem \eqref{2.6} can be
viewed as the minimal action principle of classical mechanics.  The
corresponding Hamiltonian $\bb H$ is given by
\begin{equation*}
\bb H (\rho,\pi) = \big\langle  \nabla \pi \cdot \sigma(\rho)
\nabla  \pi \big\rangle 
+  \big\langle  \nabla \pi  \cdot [ \sigma (\rho) E 
- D (\rho) \nabla \rho] \big\rangle
\end{equation*}
where at the boundary of $\Lambda$ the value of $\rho$ is prescribed
by the external reservoirs and the momentum $\pi$ vanishes
\cite{BDGJL1b}. The canonical equations associated to the Hamiltonian
$\bb H$ are
\begin{equation}
\label{f12}
\left\{
\begin{array}{l}
\rho_t + \nabla \cdot \sigma (\rho) E \;=\; 
\nabla \cdot D(\rho) \nabla  \rho
\;-\; 2 \nabla \cdot \sigma(\rho) \nabla \pi  \\
\pi_t + E \cdot \sigma'(\rho) \nabla \pi \;=\; 
- \nabla \pi \cdot \sigma' (\rho) \nabla \pi 
- D(\rho) \nabla \nabla  \pi
\end{array}
\right.
\end{equation}
in this formula, $D(\rho)\nabla \nabla \pi = \sum_{i,j} D_{i,j}(\rho)
\partial^2_{x_i, x_j} \pi$.

Recalling that $\upbar\rho$ is the stationary solution to \eqref{f1}, 
$(\upbar\rho,0)$ is an equilibrium solution of \eqref{f12}
belonging to the zero energy manifold $\bb {H}(\rho,\pi)=0$. Any
solution $\rho(t)$ of the hydrodynamical equation \eqref{f1} corresponds
to a solution $(\rho(t), 0)$ of the Hamilton equation \eqref{f12}
which converges, as $t\rightarrow +\infty$, to the
equilibrium point $(\upbar\rho,0)$ and the corresponding action
vanishes. 
The set $\{(\rho,\pi)\,:\,\pi=0\}$ is therefore the stable
manifold $\mms M_\mathrm{s}$ associated to the equilibrium position
$(\upbar{\rho}, 0)$.  
The unstable manifold $\mms M_\mathrm{u}$ is defined as the set
of points $(\rho,\pi)$ such that the solution of the canonical
equations \eqref{f12} starting from $(\rho,\pi)$ converges to
$(\upbar{\rho}, 0)$ as $t\to - \infty$.  By the conservation of the
energy, $\mms M_\mathrm{u}$ is also a subset of the zero energy manifold.

A basic result in Hamiltonian dynamics is the following \cite{arn}.
Given a closed curve $\gamma=\{ (\rho(\alpha),\pi(\alpha))
\,,\:\alpha\in [0,1]\}$, the integral $\oint_\gamma \< \pi\,d\rho\> =
\int_0^1 \langle \pi(\alpha)\, \rho_\alpha(\alpha)\rangle \, d\alpha$
is invariant under the Hamiltonian evolution. This means that, by
denoting with $\gamma(t)$ the evolution of $\gamma$ under the
Hamiltonian flow, 
$\oint_{\gamma(t)} \< \pi \,d\rho\> =\oint_\gamma \< \pi\,d\rho\>$.  
In view of this result, if
$\gamma$ is a closed curve contained in the unstable manifold $\mms
M_\mathrm{u}$ then $ \oint_\gamma \< \pi\,d\rho\>= \lim_{t\to -\infty}
\oint_{\gamma(t)} \< \pi\,d\rho\> =0$.  We can therefore define the
pre-potential $W:\,\mms M_\mathrm{u}\to\bb R$ by
\begin{equation}
\label{f23}
W(\rho, \pi) \;=\; \int_\gamma \< \hat\pi\,d\hat\rho\> \;,
\end{equation}
where the integral is carried over a path $\gamma = (\hat \rho, \hat
\pi)$ in $\mms M_\mathrm{u}$ which connects $(\upbar{\rho}, 0)$ to
$(\rho, \pi)$.  
The possibility of defining such potential is usually referred to by
saying that $\mms M_\mathrm{u}$ is a Lagrangian manifold.

The relationship between the quasi-potential and the pre-potential is
given by
\begin{equation}
\label{f13}
V(\rho) \;=\; \inf  
\big\{ W(\rho, \pi)\,,\: \pi\,: \: (\rho,\pi) \in \mms M_\mathrm{u} \big\}
\;.
\end{equation}
Indeed, fix $\rho$ and consider $\pi$ such that $(\rho,\pi)$ belongs
to $\mms M_\mathrm{u}$. Let $(\hat\rho(t), \hat\pi(t))$ be the solution
of the Hamilton equation \eqref{f12} starting from $(\rho, \pi)$ at
$t=0$.  Since $(\rho,\pi)\in \mms M_\mathrm{u}$, $(\hat\rho (t),
\hat\pi(t))$ converges to $(\bar\rho, 0)$ as $t\to -\infty$.
Therefore, the path $\hat\rho(t)$ is a solution of the Euler-Lagrange
equations for the action $I_{(-\infty, 0]}$, which means that it is a
critical path for \eqref{2.6}. 
Since 
$\bb L(\hat\rho,\hat{\rho}_t) = 
\langle \hat\pi \, \hat{\rho}_t \rangle - \bb H (\hat\rho,\hat\pi)$ and 
$\bb H (\hat\rho (t),\hat\pi(t)) =0$, 
the action of such path $\hat\rho(t)$ is given by 
$I_{(-\infty, 0]}(\hat\rho)=W(\rho,\pi)$.  The right hand side of
\eqref{f13} selects among all such paths the one with minimal
action.  

\begin{figure}[ht]
\label{f:lft}
\begin{picture}(200,110)(-100,-30) 
{
%varieta' instabile
\put(-60,55){(a)}
%
%assi
\thinlines
\put(-50,0){\line(1,0){160}}
\put(0,-30){\line(0,1){90}}
\put(-2.2,-2.2){$\bullet$}
\put(2,-8.5){$\upbar{\rho}$}
\put(112,-2.5){$\rho$}
\put(-2,63){$\pi$}
%
%varieta' instabile
\thicklines
\qbezier(-35,-15)(-10,-10)(0,0)
\qbezier(0,0)(30,30)(80,10)
\qbezier(80,10)(120,-5)(85,20)
\qbezier(85,20)(70,30)(40,30)
\qbezier(40,30)(0,30)(50,40)
\qbezier(50,40)(93,50)(95,53)
\put(-20,-11.5){\vector(-2,-1){0}}
\put(30,17){\vector(3,1){0}}
\put(70,45.5){\vector(3,1){0}}
\put(65,50){$\mms M_\mathrm{u}$}
%
%proiezione
\thinlines
\put(50,-10){\line(0,1){65}}
\put(52,-7){$\rho_\mathrm{c}$}
\put(47.7,15.5){$\bullet$}
\put(47.7,27.5){$\bullet$}
\put(47.7,37.7){$\bullet$}
%
%
%quasipotenziale
\put(140,55){(b)}
%
%assi
\thinlines
\put(150,0){\line(1,0){160}}
\put(200,-5){\line(0,1){60}}
\put(197.8,-2.2){$\bullet$}
\put(202,-8.5){$\upbar{\rho}$}
\put(312,-2.5){$\rho$}
%
%quasipotenziale
\thicklines
\qbezier(160,40)(170,0)(200,0)
\qbezier(200,0)(240,0)(250,30)
\qbezier(250,30)(280,40)(295,55)
\put(247.7,27.5){$\bullet$}
\put(165,40){$V(\rho)$}
\put(247,-7){$\rho_\mathrm{c}$}
%
%piani tangenti
\thinlines
\put(241,3){\line(1,3){14}}
\put(238,26){\line(3,1){45}}
}
\end{picture}
\caption{
  (a) Picture of the unstable manifold.
  (b) Graph of the quasi-potential. $\rho_\mathrm{c}$ is a caustic
  point. 
}
\end{figure}
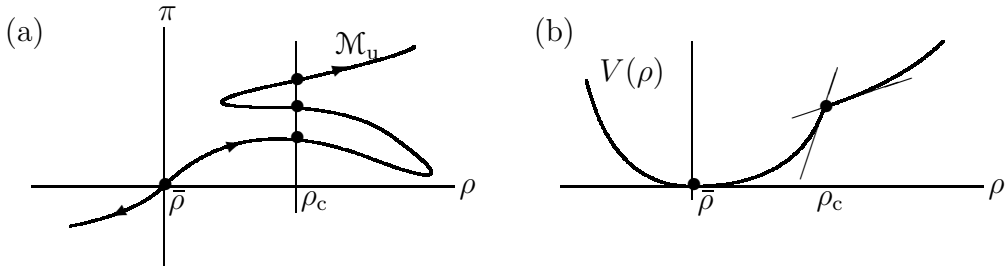

In a neighborhood of the fixed point $(\upbar{\rho},0)$, the unstable
manifold $\mms M_\mathrm{u}$ can be written as a graph, namely it has
the form $\mms M_\mathrm{u} = \{(\rho,\pi) : \pi=m_\mathrm{u}(\rho)\}$
for some map $m_\mathrm{u}$.  In this case, the infimum on the right
hand side of \eqref{f13} is trivial and $V(\rho)=
W(\rho,m_\mathrm{u}(\rho))$.  In general, though, this is not true
globally and it may happen, for special $\rho$, that the variational
problem on the right hand side of \eqref{f13} admits more than a
single minimizer (Figure 1.a).  In this case there is also more
than one minimizer for the variational problem \eqref{2.6}.  The set
of profiles $\rho$ for which the minimizer is not unique is called the
caustic. In general, it is a codimension one submanifold of the
configuration space.  We call the occurrence of this situation a
Lagrangian phase transition. In this case, profiles arbitrarily close
to each other but lying on opposite sides of the caustic are reached
by optimal paths which are not close to each other. This implies that
on the caustics the first derivative of the quasi-potential is
discontinuous (Figure 1.b). 
In particular, the occurrence of this phenomenon can be
described as a first order phase transition. Of course, there exist
also profiles for which the transition becomes of higher order.

Lagrangian phase transitions cannot occur in equilibrium.  In this
case the quasi-potential is in fact always convex, the unstable
manifold is globally a graph, and the occurrence of a first order
phase transition is due to a flat part in the quasi-potential.  In
contrast, in nonequilibrium systems the quasi-potential can be
non-convex \cite{BGL,dls3} and Lagrangian phase transitions can arise
when projecting the pre-potential $W$, which is a smooth function on
the unstable manifold $\mms M_\mathrm{u}$, onto the configuration
space.

\section{Microscopic model}

We next show that a Lagrangian phase transition occurs in a simple
nonequilibrium model, the one dimensional asymmetric simple
exclusion process on a lattice of $N$ sites with open boundaries.
Each site $i/N$, $1\le i\le N$, is either empty or
occupied by a single particle. Each particle independently attempts to
jump to its right neighboring site with rate $p$ and to
its left neighboring site at rate $q$; we assume $p>q$. 
At the boundary sites particles are added and removed: a particle is
added at site $1$, when the site is empty, at rate $\rho_0$ and
removed, when the site is occupied, at rate $1-\rho_0$; similarly
particles are added to site $N$ at rate $\rho_1$ and removed at rate
$1-\rho_1$.  
The phase diagram of the model,
corresponding to the typical behavior of
the empirical density as $N\to\infty$, 
can be derived from an algebraic 
representation of the invariant measure  \cite{dls3,san}. 
Such phase diagram
exhibits a phase coexistence when 
$0<\rho_0<\rho_1<1$ and $\rho_0+\rho_1=1$.
Note that the phase diagram can also be constructed just looking at the
entropic stationary solutions to the inviscid Burgers equation $\rho_t
+(p-q) [ \rho (1-\rho)]_x =0$ with the boundary conditions
$\rho(0)=\rho_0$, $\rho(1)=\rho_1$ \cite{BA}.

We consider the weakly asymmetric exclusion process which is obtained
by choosing $p-q= E/N$ with $E>0$ and $q=1$ \cite{de}. We also assume
$0<\rho_0<\rho_1<1$ so that there is a competition between the
external field and the boundary conditions. 
With these choices, the hydrodynamic equation, obtained in the
diffusive scaling limit, is \eqref{f1} with $\Lambda$ given by the
interval $[0,1]$, $D=1$, $\sigma = \rho (1-\rho)$, and boundary
conditions $\rho(t,0) = \rho_0$, $\rho(t,1)=\rho_1$. The unique stationary
solution of \eqref{f1}, denoted by $\upbar{\rho}_E$, can be computed
explicitly. In particular, in the weakly asymmetric regime the phase
diagram does not exhibit any phase coexistence.
The Einstein relation holds with
$s(\rho) = \rho\log \rho + (1-\rho) \log (1-\rho)$.

To compute the quasi-potential we consider the Hamiltonian flow
\eqref{f12}. It is convenient to perform the symplectic change of
variables $\varphi = s'(\rho) - \pi$, $\psi = \rho$. In the new
variables $(\varphi,\psi)$ the Hamiltonian ${\widetilde{\bb H}}
(\varphi,\psi) = \bb H (\psi, s'(\psi) -\varphi)$ reads
\begin{equation*}
{\widetilde{\bb H}} (\varphi,\psi)
= \langle \varphi_x \, \psi(1-\psi)\, \varphi_x\rangle 
- \langle [\psi_{x} + E \, \psi(1-\psi)] \, \varphi_x  \rangle
+ E\, (\rho_1 -\rho_0)
\end{equation*}
where we used that
$\rho(0)=\rho_0$, $\rho(1)=\rho_1$.
The corresponding canonical equations are
\begin{equation}
\label{nhf}
\left\{
\begin{array}{l}
\phi_t  = \phi_{xx}  - (1- 2\psi)  \, \phi_x  \, (E - \phi_x) \,,\\
\psi_t = -  \psi_{xx}  -  E\, [ \psi(1-\psi )]_x +
2\,\big[\psi(1-\psi) \, \phi_x \big]_x  \,.
\end{array}
\right.
\end{equation}

In the new variables the fixed point $(\upbar{\rho}_E,0)$ becomes
$(s'(\upbar{\rho}_E),\upbar{\rho}_E)$. The associated stable manifold
is $\{ (\varphi,\psi):\, \varphi=s'(\psi)\}$.  We claim that the
unstable manifold is given by
\begin{equation}  
\label{Sig}
\mms M_\mathrm{u} = \Big\{ (\varphi,\psi) \, :\: 
\psi = \frac{1}{1+e^\varphi} - \frac{\varphi_{xx}}
{\varphi_x (E-\varphi_x)} \,, \; 0< \varphi_x < E \Big\}\,.
\end{equation}
Indeed, 
pick a point $(\varphi,\psi)\in \mms M_\mathrm{u}$ and let $\hat\phi$ be the
solution to
\begin{equation*}
  \hat\phi_t = -\hat\phi_{xx} + \frac{1-e^{\hat\phi}}{1+e^{\hat\phi}}
  \hat\phi_x(E-\hat\phi_x) 
\end{equation*}
with initial condition $\hat\phi(0)=\varphi$. Set now 
\begin{equation*}
   \hat\psi = \frac{1}{1+e^{\hat\phi}} 
  - \frac{\hat\phi_{xx}}{\hat\phi_x(E-{\hat\phi_x})}
\end{equation*}
and observe that $\hat\psi(0)=\psi$ since $(\varphi,\psi)\in \mms M_\mathrm{u}$.
Then, a tedious computation that we omit shows that 
 $(\hat\phi, \hat\psi)$ is a solution to the canonical equations \eqref{nhf}
which converges to the fixed point
$(s'(\upbar{\rho}_E),\upbar{\rho}_E))$ as $t\to -\infty$.
Note that in the variables $(\varphi,\psi)$ the unstable manifold 
$\mms M_\mathrm{u}$ is a graph.

The computation of the pre-potential is easily achieved in the new
variables $(\varphi,\psi)$.  We start with the generating function of
the symplectic transformation. Let
$F(\rho,\varphi) = \int_0^1 \big[ s(\rho) - \rho\, \varphi \big] \, dx$
be the so-called free generating function \cite[\S~48]{arn}, so that
$\pi = \delta F/\delta\rho\,,\,\, \psi = - \delta F/\delta\varphi$.
Equivalently, $dF = \langle \pi , d\rho\rangle - \langle\psi, d\varphi\rangle$.
Hence, for any path $\Gamma=\{\gamma(t), \, t\in[0,1]\}$ in the phase
space 
\begin{equation*}
\int_\Gamma \langle \pi\, d\rho \rangle 
=  \int_\Gamma  \langle\psi \, d\phi\rangle 
+ F(\gamma(1)) - F(\gamma(0)) \;.
\end{equation*}
Assume now that $\Gamma\subset \mms M_\mathrm{u}$. By \eqref{Sig}, we have that
\begin{equation*}
\int_\Gamma  \langle\psi \, d\phi \rangle = 
\int_0^1\big[ \phi(t) - \log\big(1+e^{\phi(t)}\big) 
+ s\big(\phi_x(t) / E \big)\big]\,dx\,
\Big|_{t=0}^{t=1}\,.
\end{equation*}
Therefore, if we define $\mc G_E$ by
\begin{equation*}
\mc G_E  (\rho,\varphi) =
\int_{0}^1 \Big[ s(\rho) + s(\varphi_x/E)  
+(1-\rho)\varphi -\log\big(1+e^{\varphi}\big)  \Big] \, dx\;,
\end{equation*}
the previous identities imply that 
\begin{equation*}
\int_\Gamma \langle \pi , d\rho  \rangle 
\;=\; \mc G_E \big(\rho (1),\phi(1)\big) 
\;-\; \mc G_E \big(\rho (0),\phi(0)\big)\;.
\end{equation*}
Hence, by \eqref{f23}, $W_E(\rho,\pi) = \mc G_E (\rho,\varphi) - \mc G_E (\upbar{\rho}_E,
s'(\upbar{\rho}_E))$, where $(\varphi, \rho)\in \mms M_\mathrm{u}$.
Therefore,
\begin{equation} 
  \label{Veps}
  {V}_E (\rho) = \inf \big\{ \mc
  G_E (\rho, \varphi) \,,\: \varphi \,: \: (\varphi, \rho) \in
  \mms M_\mathrm{u} \big\} - \mc G_E (\upbar{\rho}_E,
  s'(\upbar{\rho}_E))\;.
\end{equation}
In the previous formula the condition that $(\varphi, \rho) \in
\mms M_\mathrm{u}$ can be dropped since it is equivalent to the condition that
$\phi$ is a critical point of $\mc G_E (\rho, \varphi)$. 

A similar formula for the quasi-potential ${V}_E$ in the case
$\rho_0>\rho_1$ has been obtained in \cite{de} by combinatorial
techniques.  
Analogous expression for the quasi-potential in terms of a trial
functional like $\mc G_E$ appeared in
\cite{dd,BGL,dls3,dlsprl}. However, its intrinsic significance in
terms of the Hamilton structure behind the variational problem
\eqref{2.6} is new and answers a question raised in \cite{dlsprl}. In
particular, equation (2) in \cite{dlsprl} characterizes the unstable
manifold.

\section{Lagrangian phase transitions}

We next show that, when $\rho_0<\rho_1$ and the external field 
$E$ is large enough, the weakly asymmetric exclusion process
exhibits Lagrangian phase transitions.
This is not the case when the
external field and the reservoirs push in the same direction.
We refer to \cite{BDGJL5} for the mathematical details.

We start by arguing that, when $E$ is not large, Lagrangian phase
transitions do not occur.  Let $\varphi_i = s'(\rho_i)=\log
[\rho_i/(1-\rho_i)]$, $i=0,1$, be the chemical potentials associated to
the boundary reservoirs. 
When $E=E_0=\varphi_1-\varphi_0$ there is no current and the
microscopic dynamics satisfies the detailed balance.  Therefore, in
this case, the unstable manifold is globally a graph and there exists
a unique minimizing path for \eqref{2.6}.  By perturbing around
equilibrium, this is still the case when $E$ is close to
$E_0$.

Consider now the limiting case $E=\infty$ which corresponds to the
asymmetric simple exclusion process examined in \cite{dls3}.  In this
singular limit the hydrodynamic equation \eqref{f1} becomes the
inviscid Burgers equation and shocks are possible.
In this limit the functional $\mc G_E$ becomes
\begin{equation*}
\mc G  (\rho,\varphi)  = \int_{0}^1
\Big[ s(\rho) +(1-\rho)\varphi  -\log\big(1+e^{\varphi}\big)
\Big]\,dx\;. 
\end{equation*}
Since $\mc G$ is a concave functional of $\varphi$, the minimum of
$\mc G (\rho,\varphi)$ is attained when $\varphi$ is at the boundary
of the function space. 
Since $\phi (0) =\phi_0$, $\phi(1) = \phi_1$ and $\phi$ is increasing,
the boundary of the function space is given by the step functions
$\varphi^{(y)}(x) =\varphi_0 + (\varphi_1-\varphi_0)\,\vartheta(x-y)$,
$y\in [0,1]$, where $\vartheta$ is the Heaviside function. The profile 
$\varphi^{(y)}$ jumps from $\phi_0$ to $\phi_1$ at $y$. 
The variational problem for $V$ is therefore reduced to the one
dimensional problem
\begin{equation}
\label{vf0}
\min_{y\in [0,1]} \int_{0}^1
\Big[ s(\rho) +(1-\rho)\varphi^{(y)}  -\log\big(1+e^{\varphi^{(y)}}\big)
\Big]\,dx 
\end{equation}
which is equivalent to the expression derived in \cite{dls3}.

It is not difficult to show that, if the density profile $\rho$ is
suitably chosen, \eqref{vf0} admits two minimizers. Let 
\begin{equation*}
A= 1 - \frac{ \log (1+e^{\varphi_1}) -\log (1+e^{\varphi_0})}
     {\varphi_1-\varphi_0}
\end{equation*}
and fix a density profile $\rho:[0,1]\to [0,1]$ satisfying the
following conditions, see Figure 2.  There exist $0<
y_-<y_0<y_+< 1$ such that: $\rho(y_0)=\rho(y_\pm)=A$, $\rho(x)< A$
for $x\in [0,y_-)\cup(y_0,y_+)$, $\rho(x)> A$ for
$x\in(y_-,y_0)\cup(y_+,1)$, and $\rho$ satisfies
$\int_{y_-}^{y_+}\rho\,dx = A (y_+-y_-)$. It is simple to check that
there are two global minimizers for the variational problem
\eqref{vf0}, which are given by $y_\pm$.  

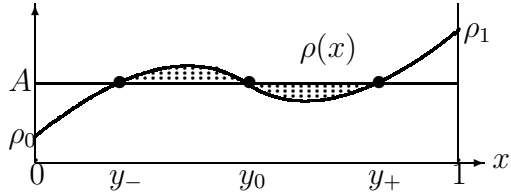
\begin{figure}[h]
\label{f:rc}
\begin{picture}(200,65)(-140,-5)
{
%
%assi
\thinlines
\put(0,0){\vector(1,0){170}}
\put(0,0){\vector(0,1){60}}
\put(-1.2,-2.2){$\cdot$}
\put(-2.2,-8){$0$}
\put(158.6,-2.2){$\cdot$}
\put(157.5,-8){$1$}
\put(160,0){\line(0,1){60}}
\put(173,-2){$x$}
%
%rho_0,rho_1, A
\put(-1.3,7.6){$\cdot$}
\put(-10,7.6){$\rho_0$}
\put(158.7,47.6){$\cdot$}
\put(162,47.6){$\rho_1$}
\put(0,30){\line(1,0){160}}
\put(-10,27){$A$}
%
%rho(x)
\thicklines
\qbezier(0,10)(50,50)(80,30)
\qbezier(80,30)(110,10)(160,50)
\put(29,27.7){$\bullet$}
\put(28,-8){$y_-$}
\put(78,27.7){$\bullet$}
\put(77,-8){$y_0$}
\put(127,27.7){$\bullet$}
\put(126,-8){$y_+$}
\put(100,40){$\rho(x)$}
%
%aree
\thinlines
\multiput(37,29)(3,0){13}{$\cdot$}
\multiput(40,31)(3,0){11}{$\cdot$}
\multiput(49,33)(3,0){7}{$\cdot$}

\multiput(84,26)(3,0){13}{$\cdot$}
\multiput(87,24)(3,0){11}{$\cdot$}
\multiput(93,22)(3,0){7}{$\cdot$}
}
\end{picture}
\caption{
  Graph of a caustic density profile for $E=\infty$.
  The shaded regions have equal area. 
}
\end{figure}

We finally argue that the occurrence of Lagrangian phase transitions
persists when the external field $E$ is large. If we
consider the density profile $\rho(x)$ in Figure 2, $\mc
G_E(\rho,\varphi)$, as a functional of $\varphi$, will have two
local minima close to $\varphi^{(y_\pm)}$ and only one of them is the
global minimizer.  However we can modify, depending on $E$, the
density profile $\rho$ in such a way that the two local minima are
brought back at the same level.  
In view of \eqref{Sig}, two optimal
paths for the variational problem \eqref{2.6} can be constructed by
the following algorithm.  Given the density profile $\rho(x)$, let
$\varphi^\pm(x)$ be two minimizers for the variational problem
\eqref{Veps} and set $F^\pm_0 = e^{\varphi^\pm}/ (1+e^{\varphi^\pm})$.
Denote by $F^\pm =F^\pm (t,x)$ the solution of the viscous Burgers
equation $F_t + E\,\big(F(1-F)\big)_x = F_{xx}$ with boundary
conditions $F(t,0)=\rho_0$, $F(t,1)=\rho_1$ and initial condition
$F(0,x)= F^\pm_0(x)$.  Set $u^\pm = s'(F^\pm)$ and define $v^\pm$ by
\begin{equation*}
v^\pm = \frac 1{1+ e^{u^\pm}}  
- \frac{u_{xx}^\pm}
{u^\pm_x (E-u^\pm_x)} 
\;\cdot
\end{equation*}
Then $v^\pm(0)=\rho$ and $v^\pm(t)$ converges to $\upbar{\rho}_E$
as $t\to +\infty$. The paths $v^\pm$ reversed in time are two
optimal paths for the variational problem \eqref{2.6}.

\section{Discussion}

We conclude with some remarks on the possibility of observing
Lagrangian phase transitions. In noisy electronic devices with a
finite number of degrees of freedom optimal paths have been
experimentally observed \cite{ML1,ML2}.  In Langevin equations with noise,
Lagrangian singularities have been observed in simulations \cite{J2,J3}.
In this paper we have shown analytically that they occur in a simple
model with infinitely many degrees of freedom. In thermodynamic
systems the thermal fluctuations are very small and the direct
observation of Lagrangian phase transitions does not appear feasible, 
as it would require an extremely long time. On the other hand, the
problem of large fluctuations admits an interpretation as a control
problem \cite{BDGJL4}. This means that rather than considering the
optimal path, we look for the field driving the system from the
stationary state to a chosen profile with the minimal energetic
cost. The Lagrangian phase transition then corresponds to the
existence of two different optimal fields dissipating the same energy.
In principle, these two fields can be theoretically calculated and an
experiment can be designed to check the predictions.

\section*{Acknowledgments}

D.G.\ acknowledges the hospitality of the Physics Department of the University of Rome La Sapienza
and the financial support of PRIN 20078XYHVYS.


\section*{References}

\begin{thebibliography}{99}

\bibitem{BDGJL1a} L. Bertini, A. De Sole, D. Gabrielli,
  G. Jona-Lasinio, C. Landim, 2001 \emph{Phys. Rev. Lett.} \textbf{87} 040601
  
\bibitem{BDGJL1b} L. Bertini, A. De Sole, D. Gabrielli,
  G. Jona-Lasinio, C. Landim, 2002 \emph{J. Statist. Phys.} \textbf{107} 635

\bibitem{BDGJL1c} L. Bertini, A. De Sole, D. Gabrielli,
  G. Jona-Lasinio, C. Landim, 2006 \emph{J. Statist. Phys.} {\bf 123} 237

\bibitem{BDGJL1d} L. Bertini, A. De Sole, D. Gabrielli,
  G. Jona-Lasinio, C. Landim, 2007 \emph{J. Stat. Mech.} P07014

\bibitem{Dcur1}
  C. Appert-Rolland, B. Derrida, V. Lecompte, F. van Wijland, 2008
  \emph{Phys. Rev. E} \textbf{78} 021122 

\bibitem{Dcur2}
  B. Derrida, A. Gerschenfeld, 2009 \emph{J. Statist. Phys.} \textbf{136} 1

\bibitem{FW} M.I. Freidlin, A.D. Wentzell, 1984 \emph{Random perturbations
  of dynamical systems}  (New York, NY: Springer-Verlag)

\bibitem{J1} R. Graham, T. T\'el, 1985 \emph{Phys. Rev. A} \textbf{31} 1109

\bibitem{J2} H.R. Jauslin, 1987 \emph{Phys. A} \textbf{144}, 179

\bibitem{J3} M.I. Dykman, M.M. Millonas, V.N. Smelyanskiy, 1994
  \emph{Phys. Lett. A} \textbf{195}, 53

\bibitem{S} H. Spohn, 1991 \emph{Large scale dynamics of interacting
  particles} (Berlin: Springer)

\bibitem{dd} B. Derrida, 2007 \emph{J. Stat. Mech.}  P07023

\bibitem{BDGJL3} L. Bertini, A. De Sole, D. Gabrielli,
  G. Jona-Lasinio, C. Landim, 2009 \emph{J. Stat.  Phys.}  {\bf 135}, 857

\bibitem{arn} V.I. Arnold,1989 \emph{Mathematical methods of classical
  mechanics} (New York, NY: Springer-Verlag)

\bibitem{BGL} L. Bertini, D. Gabrielli, J.L. Lebowitz, 2005 \emph{J. Statist. Phys.} \textbf{121} 843

\bibitem{dls3} B. Derrida, J. L. Lebowitz,  E.R. Speer, 2003 \emph{J. Statist. Phys.} \textbf{110} 775

\bibitem{san} S. Sandow, 1994 \emph{Phys. Rev. E} \textbf{50} 2660.

\bibitem{BA} C. Bahadoran, 2006 arXiv: math/0612094v2

\bibitem{de} C. Enaud, B. Derrida, 2004 \emph{J. Statist. Phys.} \textbf{114} 537

\bibitem{dlsprl} B. Derrida, J.L. Lebowitz,  E.R. Speer, 2001 \emph{Phys. Rev. Lett.} \textbf{87} 150601

\bibitem{BDGJL5} L. Bertini, A. De Sole, D. Gabrielli, G. Jona-Lasinio, C. Landim, 2010 arXiv:1004.2225

\bibitem{ML1} D.G. Luchinsky, P.V.E. McClintock, 1997 \emph{Nature} {\bf 389} 463

\bibitem{ML2} H.B. Chan, M.I. Dykman, C. Stambaugh, 2008 \emph{Phys. Rev. Lett.} \textbf{100}
130602

\bibitem{BDGJL4} L. Bertini, A. De Sole, D. Gabrielli,
  G. Jona-Lasinio, C. Landim, 2004 \emph{J. Statist. Phys.} {\bf 116}, 831

\end{thebibliography}
\end{document}